\documentclass[12pt,letterpaper,draftcls,onecolumn,journal]{IEEEtran}

\usepackage[latin1]{inputenc}
\usepackage[T1]{fontenc}
\usepackage{graphicx,varioref,subfigure,verbatim,epsfig,amsmath,mathrsfs,amsfonts,amssymb}
\usepackage[english]{babel}

\newcommand{\be}{\begin{equation}}

\newcommand{\bi}{\begin{itemize}}

\newcommand{\ee}{\end{equation}}

\newcommand{\ei}{\end{itemize}}

\newtheorem{theorem}{Theorem}
\newtheorem{lemma}{Lemma}

\date{}

\title{Joint power control and user scheduling in  multicell wireless networks: \\ Capacity scaling laws}
\author{David Gesbert (1), Marios Kountouris (1-2) \\
(1) Eurecom Institute, Sophia-Antipolis
gesbert@eurecom.fr \\
(2) France Telecom Research and Development, Issy Les Moulineaux
\thanks{Part of this work was presented at the 3rd IEEE workshop on Resource Allocation in Wireless Networks (RAWNET'07), Limassol, Cyprus, April 16th 2007.} }

\date{}

\begin{document}
\maketitle

\begin{abstract}
We address the optimization of the sum rate performance in multicell interference-limited single-hop networks where access points are allowed to cooperate in terms of  joint resource allocation. The resource allocation policies considered here combine power control and user scheduling.
 Although very promising from a conceptual point of view, the  optimization of the  sum of per-link rates hinges, in principle, on tough issues such as  computational complexity and
the requirement for heavy receiver-to-transmitter channel information feedback across all network cells.
In this paper,  we show that, in fact, distributed algorithms are actually obtainable in the asymptotic regime where the numbers of users per cell is allowed to grow large. Additionally, using extreme value theory, we provide scaling laws for upper and lower bounds for
 the network capacity (sum of single user rates over all cells), corresponding to  zero-interference and  worst-case interference scenarios. We show that the scaling is  either dominated by path loss statistics or  by small-scale fading, depending on the regime and user location scenario.

 We show that upper and lower rate bounds behave in fact identically, asymptotically. This remarkable result  suggests  not only that distributed resource allocation is practically possible but also that  the impact of  multicell interference on the capacity (in terms of scaling) actually vanishes asymptotically.
\end{abstract}
 \newpage

\section{Introduction}

The performance of wireless cellular networks is impaired by the problem of interference. Traditional ways to tackle this problem include careful planning of the spectral resource and the use of interference mitigation or advanced coding/detection techniques combined with fast link adaptation protocols at the physical layer \cite{goldsmithbook,Verdu:book}.  In a typical approach to resource planning, the system designer aims at the fragmentation of the network geographical area into smaller zones (reuse patterns), which are rendered more or less isolated from each other from a radio point of view by the assignment of slices of the spectral pie which are mutually orthogonal.  In this case, interference is reduced, but at the expense of some of the overall spectral efficiency. More subtle resource allocation protocols  include power control \cite{FOS93} and dynamic channel assignment methods which also  help alleviate the problem of interference. However the majority of such  techniques are initially rooted in voice-centric network design and, as such, are geared toward achieving a given signal to noise plus interference ratio (SINR), common to all users, rather than maximizing the spectral efficiency defined as the number of Bits/Sec/Hz per area \cite{ZAN92,YAT95}.

As an alternative to such techniques, a recently proposed approach to enhancing the network overall throughput consists in
turning interfering links into active links capable of carrying information to the user. In such as scheme the transmitters cooperate to offer a virtual MIMO setting \cite{foschini05th,shamai_2001,gesbert06co}. Although cooperation involving joint processing and encoding at all interfering transmitters constitute an efficient weapon to enhancing the system capacity, it also presents the designer with practical challenges such as limiting the overhead of cell-to-cell communication and the need for symbol-accurate synchronization between the transmitters.

Another important limitation common to most of  the above mentioned work is the fact that interference mitigation and joint transmitter processing are addressed, as physical layer enhancements,  while ignoring the possible impact of the user scheduling algorithm implemented in the multiple access protocol. We already know that single-cell greedy user scheduling has a profound impact on the fading statistics seen by the above layers of the system. In the case of multicell user scheduling, we may naturally wonder if interference may be impacted as well. In other words, a broad question is what is the impact of interference when a scheduling protocol is available that allows us to select within a pool of users?

Thus in this paper, we revisit the problem of interference in single-hop cellular wireless networks and examine it in the light of multicell resource allocation algorithms. The considered algorithms
involve some level of cooperation between the transmitters in the form of joint power control and user scheduling, with the aim
of  maximizing of the network throughput. Since joint processing/encoding is not considered here, in a first approach, the network throughput is defined in an   information theoretic sense as the sum over all cells of the per-cell single-user decoding rates, achieved by simultaneously receiving users.

A simple intuitive idea behind multicell resource allocation is to exploit the large amount of spatial and multiuser diversity offered by the extra multicell dimension in order to optimize the network performance at all times. Clearly
the potential gains comes with great challenges. One is the complexity associated with the joint optimization
of a large number of parameters (slot assignment, power levels, ..). Another one is the need for the joint processing of multicell channel state information   which necessitates   huge cell-to-cell signaling overhead. This  makes global network coordination hard to realize, especially in fast mobile settings. Despite the challenges, some recently published work suggests possible techniques for low complexity and distributed resource allocation. Examples of such approaches include game theoretic algorithms with pricing \cite{SAR02,ALT04}, resource allocation based on quantized power levels, and iterative/greedy capacity maximization techniques. An overview of such techniques is available in \cite{gesbert07ad}. Nonetheless it remains that such approaches are suboptimal and do not help characterize the system behavior when the number of user is large.

In this paper we address the problem of sum rate-maximizing resource allocation, where the protocol takes  the form of joint multicell power control and scheduling. We investigate upper and lower bounds on the maximum network capacity provided by resource allocation in interference-free and full-powered interference scenarios, respectively.
Interestingly the solution to the multicell scheduling and power control is fully distributed in both scenarios. We study these bounds in the asymptotic regime where the number of users per cell is allowed to grow large while the number of cells remain fixed. We introduce scaling laws of capacity for this asymptotic regime, based on extreme value theory. We show that for certain idealized network settings, our models coincide mathematically with recently published models in the different context of opportunistic single cell beamforming, in particular those of e.g. \cite{Viswanath02op,Sharif_RBF}. Therefore certain results from this domain can be reused. For other general network settings, different results of extreme value theory must be explored.

After identifying the scaling laws for the upper and lower bounds on capacity for two different forms of network/channel models, we draw the surprising conclusion that these are in fact identical, in both cases. In the particular case of users located at random in a finite disk around each access point, the analysis reveals a much higher growth rate than in the case of users having an equal average SNR.

One of the practical impacts  of our results is the suggestion that very simple distributed resource allocation algorithms can be used with little loss of performance on the network capacity compared with a centralized optimal scheme. Another interesting lesson is a better understanding  of the the price paid due to multicell interference
in terms of rates,  when the number of users becomes large and when a maximum-rate scheduler is exploited.
Our theoretical claims are backed with Monte-Carlo simulations on multicell fading channels.

 \section{Network and signal models} \label{sec:model}

We consider  a wireless network featuring a number of transmitters and receivers. Among these, there are $N$ transmit-receive active pairs, which are simultaneously selected for transmission by the scheduling protocol at any considered instant of time, others remaining silent. All active links interfere with each other.  This setup, an instance of the interference channel \cite{Cover91} can be observed in e.g. a  cellular network with reuse factor one, such as the upcoming 802.16 (WiMax) and 3GPP (LTE) wireless standards.
 We  assume each  of the $N$ cells is equipped with an access point (AP) and that APs communicate with the users in a single-hop fashion. We also assume the APs are time-synchronized. In this paper we focus on the performance of downlink communication from the AP to the users. However we believe our analysis carries over to the uplink without great difficulty.

   Let $U_n$ be the number of users randomly distributed over cell $n$, for $n=1,\hdots,N$. We will assume these users are uniformly randomly distributed  over either a circle or a  disk around their access point.

\begin{figure}[h]
\centering
\includegraphics[scale=0.45,angle=270]{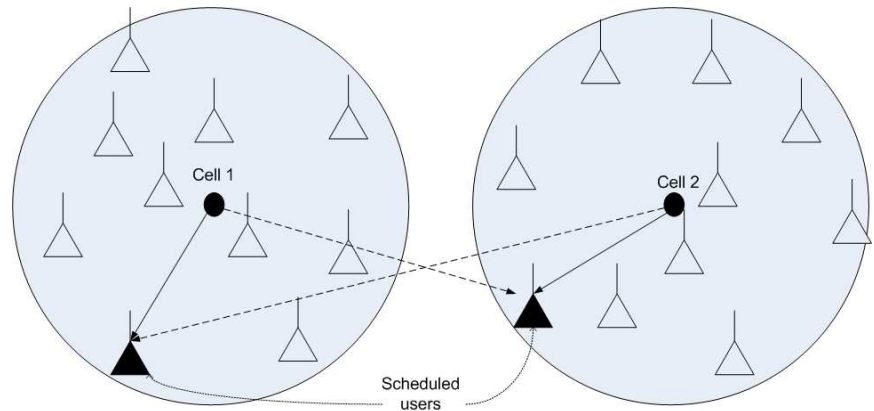}
\caption{A two-cell network diagram example. Direct and interfering links toward the scheduled user (black) are indicated in solid and dashed arrows respectively. Users are located randomly over a cell of radius $R$ around their access point.}
\label{fig:asym2cells}
\end{figure}

  Within  each cell, we consider an orthogonal multiple access scheme so that a {\em single} user is supported on any given spectral resource slot. A resource slot can be a time or frequency slot in TDMA/FDMA, or a code in orthogonal CDMA. For instance in OFDMA-based WiMax or LTE standards, a resource slot is represented by a unique time/frequency slice. For ease of exposition, single antenna devices are considered. Generalizations to MIMO-aided spatial division multiple access are currently under investigation. On any given spectral resource slot, shared by all $N$ cells, we denote by  $u_n \in \{ 1,\hdots,U_n \}$  the index of the user that is granted access to the slot (i.e. scheduled) in cell $n$.
 An example of such a situation is depicted for a simple two cell network in Fig.\ref{fig:asym2cells}.

 We denote the complex downlink channel gain between the $i$-th AP  and user $u_n$ of cell $n$ by $\alpha_{u_n, i}$.  The local channel state information (CSI) is assumed perfect at the receiver side. This information is also fedback to the control unit responsible for resource allocation, either in a centralized or distributed manner (this point crucial when it comes to applicability, as discussed   later).  The received signal $Y_{u_n}$ at  user $u_n$ is  given by
\begin{displaymath}
Y_{u_n} = \alpha_{u_n,n} X_{u_n} + \sum_{i\neq n}^{N} \alpha_{u_n,i} X_{u_i} + Z_{u_n},
\end{displaymath}
where $X_{u_n}$ is the message-carrying signal from the serving AP, subject to a peak power constraint $P_{max}$. $\sum_{i\neq n}^{N} \alpha_{u_n,i}X_{u_i}$ is the sum of interfering signals from other cells and $Z_{u_n}$ is the additive noise or extra interference.  $Z_{u_n}$ is modeled for convenience as white Gaussian with power $\mathbb{E}|Z_{u_n}|^2 = \sigma^2$.
Note that a single power level is applied at each AP in this notation. This will allow us to ease the exposition of our analysis. In the OFDMA case however, a possibly unequal power level may be applied on each subcarrier, leading to the optimization of a {\em power vector}, under sum power constraint,  rather than a {\em scalar} power level at each AP. The analysis in that case however leads to similar conclusions on the capacity scaling and is skipped in this paper.

 \section{The multicell resource allocation problem}  \label{sec:models}
As stated above, intra-cell multiple access is orthogonal, while intercell multiple access is simply superposed, due to full reuse of spectrum.  The resource allocation problem considered here consists in  {\em power allocation} and {\em user scheduling} subproblems. Importantly we focus on {\em rate maximizing} resource allocation policies, rather than {\em fairness-oriented} ones. As is the case with known single cell protocols, multicell scheduling protocols can be enhanced
to offer some desired performance-fairness trade-off, however this is outside the focus of this paper. Fairness issues are touched upon in \cite{gesbert07ad}. In our setting the optimization of resource in the various resource slots decouples and we can consider the power allocation and user scheduling maximizing the capacity in any one slot, independently of other slots.  A few useful definitions follow.

\newtheorem{define}{Definition}
\begin{define}\label{schvec}
A \emph{\textbf{scheduling vector}} $\boldsymbol{U}$ contains the set of users simultaneously scheduled across all $N$ cells in the same slot:
\begin{displaymath}
\boldsymbol{U} = [u_1 \, u_2 \, \cdots \, u_n \, \cdots \, u_N]
\end{displaymath}
where $[\boldsymbol{U}]_n=u_n$. Noting that $1 \leq u_n \leq U_n$, the constraint set of scheduling vectors is given by $ \Upsilon = \{\boldsymbol{U} \mid  1 \leq u_n \leq U_n \ \forall \ n = 1, \dots,N\}$.
\end{define}
\begin{define}
A \emph{\textbf{transmit power vector}} $\boldsymbol{P}$ contains the transmit power values used by each AP to communicate with its respective user:
\begin{displaymath}
\boldsymbol{P} = [P_{u_1} \, P_{u_2} \, \cdots \, P_{u_n} \, \cdots \, P_{u_N}]
\end{displaymath}
where $[\boldsymbol{P}]_n=P_{u_n}=\mathbb{E}|X_{u_n}|^2$. Due to the peak power constraint $0 \leq P_{u_n} \leq P_{\max}$, the constraint set of transmit power vectors is given by $ \Omega = \{\boldsymbol{P} \mid  0 \leq P_{u_n} \leq P_{\max}\ \forall \ n = 1, \dots,N \}$.
\end{define}

 \subsection{Capacity optimal resource allocation}
  The merit (or utility) associated with a particular choice of a scheduling vector and power allocation vector is  measured via the set of SINRs observed by all scheduled users simultaneously.
    $\Gamma([\boldsymbol{U}]_n,\boldsymbol{P})$ refers to the SINR experienced by the receiver $u_n$ in cell $n$ as a result of power allocation in all cells, and is  given by:
    \begin{equation}
\Gamma([\boldsymbol{U}]_n,\boldsymbol{P}) = \frac{G_{u_n, n}P_{u_n}}{\displaystyle \sigma^2 + \sum_{i\neq n}^{N}G_{u_n, i}P_{u_i}},
\label{eq:gam}
\end{equation}
  where   $G_{u_n, i}=|\alpha_{u_n, i} |^2$ is the channel power gain from cell $i$ to receiver $u_n$.

   In data-centric applications, a reasonable choice of utility is a monotonically piece-wise increasing function of the SINR, reflecting the various coding rates implemented in the system. With an idealized link adaptation protocol, the utility eventually converges to a smooth function reflecting the user's instantaneous rate in Bits/Sec/Hz. For the overall network utility, and under the assumption that the various cell transmitters cannot afford to perform cooperative coding,  we consider the average of rates achieved over all cells under single-user decoding, thus absorbing all sources of interference   under a global Gaussian interference  process. We may write the utility as \cite{Cover91}:

\begin{equation} \label{eq:objfunction}
\mathcal{C}(\boldsymbol{U},\boldsymbol{P}) \stackrel{\Delta}{=} \frac{1}{N}\sum_{n=1}^{N}\log\Big(1+\Gamma([\boldsymbol{U}]_n,\boldsymbol{P})\Big).
\end{equation}

The capacity optimal resource allocation problem can now be formalized simply as:

\begin{equation}
(\boldsymbol{U}^*,\boldsymbol{P}^*) = \mathrm{arg}\max_{\substack{\boldsymbol{U} \in \Upsilon \\ \boldsymbol{P} \in \Omega }} \mathcal{C}(\boldsymbol{U},\boldsymbol{P}),
\label{eq:optimal}
\end{equation}
The optimization above can be seen as generalizing known approaches in two ways. First the capacity maximizing scheduling problem has been considered (e.g. \cite{KNOPP95}), but in general not jointly over multiple cells. Second, the problem above extends the classical multicell power control problem (which usually rather aims at achieving SINR balancing) to include joint optimization with the scheduler.

The problem in   (\ref{eq:optimal}) presents the system designer with many degrees of freedom to boost system capacity but also with several serious challenges.
First the problem above is non convex and standard optimization techniques do not apply directly. On the other hand an exhaustive search of the $(\boldsymbol{U},\boldsymbol{P})$ pairs over the constraint set is prohibitive. Finally, even if computational issues were to be resolved, the optimal solution still requires a central controller updated with instantaneous inter-cell channel gains which would create acute signaling overhead issues in practice. The central question addressed by this paper can be formulated as follows: Can we extract all/some of the gain related to multicell resource allocation (compared with single cell treatment) within reasonable complexity and signaling constraints? Inspection of the recent literature reveals that this is a hot research issue with many possible tracks of investigation including use of modified capacity metrics, game theoretic approaches,  reuse partitioning, power shaping and power quantizing  (see e.g. \cite{gesbert07ad} and references therein).  Below, we investigate theoretical answers on this question by means of so-called {\em scaling laws} of the capacity, obtained via extreme value theory. This study reveals both surprising and promising answers.

Interestingly, other work exists on analyzing the scaling law of capacity in  interference-limited networks, including  recently submitted \cite{ebrahimi}. In such work, a similar metric is used related to the sum capacity of simultaneously active links. Importantly though, in their work the scaling is in terms of growing number of links (or cells for a cellular network), rather than growing number of users per cell in a multiple access scheme as is studied here. Thus multiuser diversity scheduling  is not exploited and fundamentally different scaling laws are obtained in the two cases.

 \section{Network capacity: Models and bounds} \label{sec:asymp}
Let us consider a system with a large number of users in each cell. For the sake of exposition we shall assume $U_n=U$ for all $n$, where $U$ is asymptotically large, while $N$ remains fixed. We expect a growth of the sum capacity $\mathcal{C}(\boldsymbol{U}^*,\boldsymbol{P}^*)$ with $U$ thanks to the {\em multicell} multiuser diversity gain\footnote{The multicell multiuser diversity gain can be seen as a generalization of the conventional multiuser diversity \cite{KNOPP95} to multicell scenarios with joint scheduling}. Thus we are interested in how the {\em expected} sum capacity  {\em scales} with $U$. To this end we shall use several interpretable bounding arguments. We consider two channel gain models. The first considers a symmetric distribution of gains to all users from their serving AP. In the other one, an additional random distance-dependent path loss is accounted for.

\subsection{Bounds on network capacity}

The simple bounds below hold in the asymptotic and non asymptotic regimes as well.

{\bf Upper bound:} An upper bound (ub) on the capacity for a given resource allocation vector (not necessarily an optimal one) is obtained by simply ignoring intercell interference effects:

\begin{equation}
\mathcal{C}(\boldsymbol{U},\boldsymbol{P}) \leq  \frac{1}{N}\sum_{n=1}^{N}\log\Big(1+  \frac{G_{u_n, n}P_{u_n}}{\displaystyle \sigma^2 }   \Big).
\end{equation}
In the absence of interference, the optimal capacity is clearly reached by transmitting at a level equal to the power constraint, i.e.  ${\bf P}_{max}=[P_{max},\hdots,P_{max}]$ and selecting the user with the largest channel gain in each cell (maximum rate scheduler), thus giving the following upper bound on capacity:

\begin{equation}
\mathcal{C}(\boldsymbol{U}^*,\boldsymbol{P}^*) \leq \mathcal{C}^{ub}
\end{equation}

where

\begin{equation}
\mathcal{C}^{ub} = \frac{1}{N}\sum_{n=1}^{N}\log\Big(1+
\Gamma_n^{ub}  \Big).
\label{eq:upperbound}
\end{equation}

and where the upper bound on SINR $\Gamma_n^{ub}$ is given by the maximum rate scheduler:

\begin{equation}
 \Gamma_n^{ub}=   \max_{u_n=1\hdots  U} \{ G_{u_n, n}\} P_{max}/\sigma^2
\label{eq:snroptimal}
\end{equation}

{\bf Lower bound:}
A lower bound (lb) on the optimal  capacity (in the presence of interference) $\mathcal{C}(\boldsymbol{U}^*,\boldsymbol{P}^*)$ can be derived by restricting the domain of optimization. Namely, by restricting the power allocation vector to full power $P_{max}$ in all transmitters,  we have
\begin{equation}
\mathcal{C}(\boldsymbol{U}^*,\boldsymbol{P}^*) \geq \mathcal{C}^{lb}
\end{equation}

where

\begin{equation}
\mathcal{C}^{lb} = \mathcal{C}(\boldsymbol{U}^*_{FP}, {\bf P}_{max}) \label{eq:lowerbound}
\end{equation}

and where  $\boldsymbol{U}^*_{FP}$ denotes the maximum rate scheduling vector when assuming full power everywhere. This scheduling vector is defined  by

\begin{equation}
\boldsymbol{U}^*_{FP} = \mathrm{arg}\max_{\substack{\boldsymbol{U} \in \Upsilon }} \mathcal{C}(\boldsymbol{U},\boldsymbol{P}_{max}),
\label{eq:SINRoptimal}
\end{equation}

Note that the user selected in the $n$-th cell, designated by $[\boldsymbol{U}^*_{FP}]_n$, is found via:

\begin{equation}
[\boldsymbol{U}^*_{FP}]_n = \mathrm{arg}\max_{\substack{\boldsymbol{U} \in \Upsilon }}
  \frac{\{ G_{u_n, n}\} P_{max}}{\sigma^2+\sum_{i\neq n}^{N}G_{u_n, i}P_{max}}
\label{eq:SINRoptimal2}
\end{equation}

 The  SINR corresponding to the selected user, denoted by $\Gamma_n^{lb}$, is therefore  given by:

\begin{equation}
 \Gamma_n^{lb}=   \max_{u_n=1\hdots  U}  \frac{\{ G_{u_n, n}\} P_{max}}{\sigma^2+\sum_{i\neq n}^{N}G_{u_n, i}P_{max}}
\label{eq:SINRoptimal3}
\end{equation}

Finally the lower bound on capacity $\mathcal{C}^{lb}$ may be rewritten as:

\begin{equation}
 \mathcal{C}^{lb}=   \frac{1}{N}\sum_{n=1}^{N}\log\Big(1+
\Gamma_n^{lb}  \Big).
\label{eq:lowerboundcapa}
\end{equation}

\subsection{Distributed vs. centralized scheduling}

  For large networks, it is important that scheduling algorithms can operate on a distributed mode, that is, the choice of the
  optimal user set should  be done by each cell on the basis of locally available information only. This is in principle difficult task because  the achievable rates observed in different cells are coupled together through the interference terms. However, we note that if the scheduler is based on maximizing the upper bound of network capacity given by (\ref{eq:upperbound}), then each cell only needs to know the realization of the direct gain $G_{u_n, n}$, and the scheduler is trivially distributed. Alternatively, to obtain a scheduler  maximizing the lower bound of capacity given by (\ref{eq:lowerboundcapa}), each cell must collect the worst case SINR for each of its users.
  The worst case SINRs are computed during e.g. a common preamble phase where all APs are asked to transmit pilot or data symbols at full power. This makes the scheduler of (\ref{eq:SINRoptimal2}) also distributed.  Note that  "worst case" is here understood in terms relative to the power control policy, not the scheduler.

\subsection{Channel models}
 We now detail our assumptions regarding the fading and path loss models. Some of these assumptions are mainly  technical, serving to simplify the analysis but could be relaxed without altering the fundamental results, as discussed later. As mentioned above we assume a cellular network where APs are regularly located with cell radius $R$. In this sense, the cells are assumed to be circular with each base being at the center of it, although this assumption is not critical to this study (i.e. similar conclusions can be obtained for hexagonal cell etc.) as explained below.

The basic channel model consists in the product between a variable representing the path loss  and a variable representing the fast fading coefficient: Let $G_{u_n,i}=\gamma_{u_n,i} | h_{u_n,i}|^2$, $u_n=1\hdots  U, i=1\hdots  N$ be the set of power gains where $\gamma_{u_n,i}$ is the path loss between user $u_n$ (selected in cell $n$) and the access point in cell $i$.  $ h_{u_n,i}$ is the corresponding normalized complex fading coefficient.
A generic path loss model is given by \cite{lee82mo}:

\begin{equation}
\gamma_{u_n,i}= \beta d_{u_n,i}^{-\epsilon}
\label{eq:pathl}
\end{equation}
where $\beta$ is scaling factor, $\epsilon$ is the path loss exponent (usually with $\epsilon>2$), and $d_{u_n,i}$ is the distance between user $u_n$ and AP $i$.

Note that we assume  as preamble a user-to-AP assignment strategy resulting in all users being served by the AP with the smallest path loss. This means, as is usually the case in current network design,  that the AP assignment operates on a time scale which is not fast enough to provide diversity against fast fading.

\begin{figure}[h]
\centering
\includegraphics[scale=0.45,angle=270]{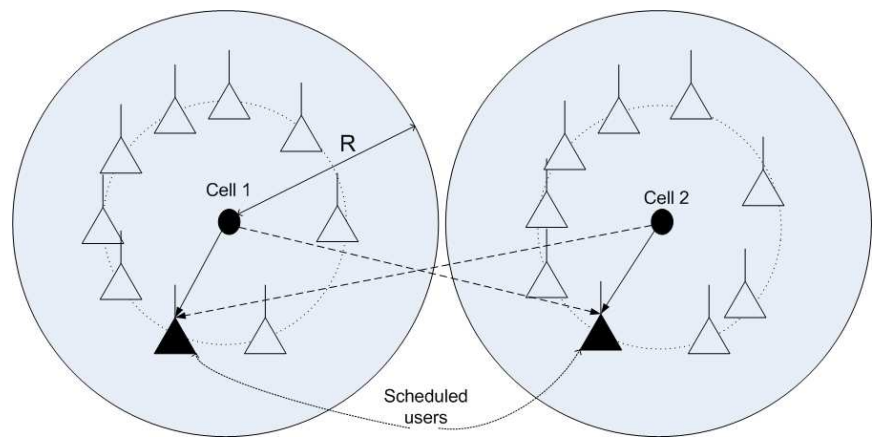}
\caption{A two-cell idealized symmetric network diagram example. Direct and interfering links toward the scheduled user (black) are indicated in solid and dashed arrows respectively. In this idealized case, users a located over a circle, a fixed distance away from  their access point.}
\label{fig:sym2cells}
\end{figure}

We consider in turn two basic user location scenarios, and a hybrid one. As it will be made clear later,
the user location scenario has significant impact on the analysis of the network capacity. In the first scenario, denoted as {\em symmetric network}, all users served by a given AP are assumed to be located at the same distance from that AP. This idealized situation results in all users experiencing the same average SNR, an assumption often made by previous authors in this area, and for which several interesting results of the existing literature can be reused.  This scenario is illustrated in Fig.\ref{fig:sym2cells}.

In the second, more realistic, scenario, denoted simply as {\em non-symmetric network}, the users are located randomly over a cell given by a disk of radius $R$ around each of the serving APs. Finally, a hybrid scenario mixing the two scenarios above is discussed later in the paper.

  Note that the actual cell shape will not be a disk in reality. However we argue that, when it comes to studying the scaling laws of network capacity with maximum-capacity user scheduling, the actual shape taken by  the cell borders has in fact little impact on the result. The main reason is that since
the user's direct links is subject to a location dependent path loss,  the distance to the serving AP  will affect its chances of being selected by the scheduler. As a consequence the users located in the inner region of the cell (i.e. close to the access point)  bear the vast majority of the traffic and are the drivers for the capacity scaling laws. Therefore an accurate modeling for the location of cell-edge users is unimportant here.

 \section{Network capacity: Scaling laws} \label{sec:asymp2}

\subsection{Capacity scaling with large $U$ in symmetric network}
\label{sec:symm}
We analyze the scaling of capacity  $\mathcal{C}(\boldsymbol{U}^*,\boldsymbol{P}^*)$ via the scaling of the bounds $\mathcal{C}^{lb}$ and $\mathcal{C}^{ub}$, with increasing $U$. We just focus on the performance in cell $n$, as other cells are expected to behave similarly under equal number of users $U$ and isotropic conditions throughout the network.
For the symmetric network, users experience an equal average SNR, thus $\gamma_{u_n, n}=\gamma_n$ is a constant independent of the user index.

Interestingly, for this particular case, we show we can reuse extreme value theory results \cite{david03} developed specifically in the context of single cell opportunistic  beamforming \cite{Viswanath02op,Sharif_RBF} and transposed here to the case of networks with multicell interference. For the case asymmetric network, specific results are developed.

  First, the following results provide insight into the interference-free  scaling of SINR and capacity respectively.

\subsubsection{Scaling laws for interference-free case}
The interference-free multicell capacity scaling boils down to studying the scaling in each cell independently. Further assuming an isotropic network (i.e. all cells experience the same channel statistics) we can simplify the analysis by exploiting known results on single-cell capacity scaling, as done below.

 \begin{lemma}
 Let $G_{u_n,n}=\gamma_{u_n,n} | h_{u_n,n}|^2$, $u_n=1\hdots U, n=1\hdots N$, where $\gamma_{u_n,n}=\gamma_n$. Assume $| h_{u_n,n}|^2$ is Chi-square distributed with 2 degrees of freedom ($\chi^2(2)$) (i.e. $h_{u_n,n}$
 is a unit-variance complex normal random variable). Assume the $| h_{u_n,n}|^2$ are independent and identically distributed (i.i.d.) across users.  Then for fixed $N$ and $U$ asymptotically large, the upper bound on the SINR in cell $n$ scales like
\begin{equation}
\label{eq:snrcaling}
\Gamma_n^{ub} \approx  \frac{P_{max}\gamma_n}{\sigma^2}\log U
\end{equation}
where the symbol $\approx$ means that the ratio of the left hand side and right hand side terms converges to one, as $U$ goes to infinity.
\end{lemma}

{\bf Proof:} This result is a reuse of a now well known result for single cell opportunistic scheduling. This states  that the maximum of $U$ i.i.d. $\chi^2(2)$ random variables behaves like $\log U$ for large $U$. See for instance \cite{Viswanath02op}, itself building on classical extreme value theory results \cite{david03}. We  omit the proof here and refer the readers to these references.

\vspace*{0.6cm}

From the scaling the SNR, we obtain the scaling of the interference-free capacity shown in \ref{eq:upperbound}. This is stated in the following theorem, again building on known single cell results but stated here for convenience, with our own notations:

 \begin{theorem} \label{theorem1}
 Let $G_{u_n,n}=\gamma_{u_n,n} | h_{u_n,n}|^2$, $u_n=1\hdots U, n=1\hdots N$, where $\gamma_{u_n,n}=\gamma$. This means that all cells are assumed to enjoy an identical link budget. Assume $| h_{u_n,n}|^2$ is Chi-square distributed with 2 degrees of freedom ($\chi^2(2)$). Assume the $| h_{u_n,n}|^2$ are i.i.d. across users.  Then for fixed $N$ and $U$ asymptotically large, the average of the upper bound on the  network capacity scales like
\begin{equation}
\label{eq:capascaling1}
\mathbb{E}(\mathcal{C}^{ub}) \approx \log \log U
\end{equation}
where the expectation is taken over the complex fading gains.
\end{theorem}

  {\bf Proof:} Under isotropic network conditions, we have from (\ref{eq:upperbound}):
  \begin{equation}
\mathbb{E}(\mathcal{C}^{ub}) = \mathbb{E}\left( \log\Big(1+
\Gamma_n^{ub}  \Big)\right)
\end{equation}

  Once the scaling of $\Gamma_n^{ub}$ is obtained, the scaling of the expected value of $\log (1+
\Gamma_n^{ub}  )$ is readily obtained from published results in the context of single-cell maximum rate user scheduling, found  in  \cite{Viswanath02op,Sharif_RBF} among others. For a detailed proof, see e.g. \cite{Sharif_RBF}, Theorem 1.

  \vspace*{0.6cm}

\subsubsection{Scaling laws for full-powered interference case}
We now turn to the behavior of interference limited networks by exploring the lower bounds given for SINR and capacity.
The initial intuition would be that the analysis of the lower bound given in (\ref{eq:SINRoptimal3}) provides us with a  very pessimistic view of the network performance as it assumes  interference coming at full power from every AP in the network. The interesting aspect behind our findings below is that it is not. In fact the negative impact of interference at the user on network capacity can be made arbitrarily small while not sacrificing transmission rates to the assigned APs, as shown per the following theorems.
In the results below, remember we assume each user is assigned to a serving AP which is the one with minimum path loss. As a consequence, since the region of coverage under study is limited to a disk of radius $R$ around the serving AP, the distance between a user and any interfering AP is greater than $R$. As a result we have from (\ref{eq:pathl}):

\begin{equation}
G_{u_n,i}\leq \beta R^{-\epsilon} |h_{u_n, i}|^2  \ \ \ \mbox{for any} \ i \neq n
\label{eq:distance_upperbound}
\end{equation}

The lemma below gives the scaling law for the worst case SINR \ref{eq:SINRoptimal3}.

 \begin{lemma} \label{lemma2}
 Let $G_{u_n,i}=\gamma_{u_n,i} | h_{u_n,i}|^2$, $u_n=1\hdots U, n=1\hdots N$, where $\gamma_{u_n,n}=\gamma_n$, $\gamma_{u_n,i}=\beta d_{u_n,i}^{-\epsilon}$ for $i\neq n$. Assume $| h_{u_n,i}|^2$ is Chi-square distributed with 2 degrees of freedom ($\chi^2(2)$). Assume the $| h_{u_n,i}|^2$ are i.i.d. across users, cells. Then for fixed $N$ and $U$ asymptotically large, the lower bound on the SINR in cell $n$ scales like
\begin{equation}
\label{eq:sinrcaling}
\Gamma_n^{lb} \approx  \frac{P_{max}\gamma_n}{\sigma^2}\log U
\end{equation}
\end{lemma}

{\bf Proof:} To obtain this result, one uses the fact that users in cell $n$ are served by their closest AP. Following (\ref{eq:distance_upperbound}), an upper bound on the interference power then given by
$\sum_{i\neq n}^{N}\beta R^{-\epsilon} |h_{u_n, i}|^2 P_{max}$. This gives a further lower bound on $\Gamma_n^{lb}$ given by

\begin{equation}
 \Gamma_n^{lb} \geq  \Gamma_n^{lb2}
 \end{equation}
 where $\Gamma_n^{lb2}$ is corresponds to the SINR assuming pessimistically that all sources of interferences are located on the edge of the cell of interest, calculated by:
 \begin{equation}
\Gamma_n^{lb2}  = \gamma_{n} P_{max} \max_{u_n=1\hdots U}
  \omega_{u_n}
\label{eq:SINRoptimal4}
\end{equation}
where $\omega_{u_n}$ denotes the normalized SINR at user $u_n$:
\begin{equation}
 \omega_{u_n}= \frac{| h_{u_n,n}|^2 }{\sigma^2/P_{max}+ \beta R^{-\epsilon} \sum_{i\neq n}^{N} |h_{u_n, i}|^2}
  \label{eq:omeg}
  \end{equation}
The scaling law of $\Gamma_n^{lb2}$ is also that of
$\omega_{u_n}$, which is the ratio of a Chi-square (2 degrees of freedom) distributed variable and the sum of a fixed noise term and a Chi-square (2N-2 degrees of freedom) variable.
 Thus the scaling of  $\omega_{u_n}$ is similar to the scaling
of the SINR in the single cell opportunistic beamforming problem with $N$ antennas at the transmitter, studied in \cite{Sharif_RBF}. In there, the SINR is the ratio of a direct beam power and a noise plus $N-1$ interfering beam power term. In particular we can find its distribution as:

\be
F_W(\omega)= 1 - \frac{e^{-\frac{\omega \sigma^2}{P_{max}}}}{(1+\omega \beta (N-1)R^{-\epsilon})^{N-1}}
\ee

(\cite{Sharif_RBF}, Lemma 4) shows that the SINR then scales like $\log U$. This gives in our context:
 \be
  \Gamma_n^{lb2} \approx P_{max}\gamma_n \log U /\sigma^2
  \ee
  Note that the scaling above is identical to the one reported for the interference-free case (\ref{eq:snrcaling}).

  Thus, $\Gamma_n^{lb}$ is bounded above and below by two expressions (respectively the interference-free $\Gamma_n^{ub}$ and $\Gamma_n^{lb2}$) which exhibit the same scaling law. Therefore $\Gamma_n^{lb}$ must satisfy itself the same scaling law.

\vspace*{0.6cm}

The following theorem gives the scaling law for the lower bound on capacity for an isotropic network.

 \begin{theorem} \label{theorem2}
 Let $G_{u_n,i}=\gamma_{u_n,i} | h_{u_n,i}|^2$, $u_n=1\hdots U, n=1\hdots N$, where $\gamma_{u_n,n}=\gamma$, $\gamma_{u_n,i}=\beta d_{u_n,i}^{-\epsilon}$ for $i\neq n$. Assume $| h_{u_n,i}|^2$ is Chi-square distributed with 2 degrees of freedom ($\chi^2(2)$). Assume the $| h_{u_n,i}|^2$ are i.i.d. across users, cells.  Then for fixed $N$ and $U$ asymptotically large, the average of the lower bound on the  network capacity scales like
\begin{equation}
\label{eq:capascaling2}
\mathbb{E}(\mathcal{C}^{lb}) \approx \log \log U
\end{equation}
\end{theorem}

  {\bf Proof:} From the result in Lemma \ref{lemma2},  this result is proved in a way identical with that of (\cite{Sharif_RBF}, Theorem 1). Therefore the proof is omitted here for space considerations.

  \vspace*{0.6cm}

From bounding arguments and from theorems \ref{theorem1} and \ref{theorem2} above, the following conclusion is now obtained:

 \begin{theorem} \label{theorem21}
 Let $G_{u_n,i}=\gamma_{u_n,i} | h_{u_n,i}|^2$, $u_n=1\hdots U, n=1\hdots N$, where $\gamma_{u_n,n}=\gamma_n$, $\gamma_{u_n,i}=\beta d_{u_n,i}^{-\epsilon}$ for $i\neq n$. Assume $| h_{u_n,i}|^2$ is Chi-square distributed with 2 degrees of freedom ($\chi^2(2)$). Assume the $| h_{u_n,i}|^2$ are i.i.d. across users, cells.  Then for fixed $N$ and $U$ asymptotically large, the average of the  network capacity with optimum power control and scheduling scales like
\begin{equation}
\label{eq:capascaling22}
\mathbb{E}(\mathcal{C}(\boldsymbol{U}^*,\boldsymbol{P}^*) ) \approx \log \log U
\end{equation}
\end{theorem}

 {\bf Proof:} The result is readily obtained from writing:
\be
\mathbb{E}(\mathcal{C}^{lb}) \leq \mathbb{E}(\mathcal{C}(\boldsymbol{U}^*,\boldsymbol{P}^*)) \leq \mathbb{E}(\mathcal{C}^{ub})
\ee
 Then, invoking (\ref{eq:capascaling2}) and (\ref{eq:capascaling1}) exhibiting the same scaling law, we obtain a similar law in
(\ref{eq:capascaling22}).

\vspace*{0.6cm}

Theorems \ref{theorem1} and \ref{theorem2} suggest that, in a multicell network with symmetric users, the capacity obtained with optimal multicell scheduling in both an interference-free environment and an environment with full interference power have identical scaling laws in $\log \log U$. This result bears analogy to the results by \cite{Sharif_RBF} which indicate that in a single cell broadcast channel with random beamforming and opportunistic scheduling, the degradation caused by inter-beam interference tends becomes negligible when the number of users to choose from becomes large. Here the multicell interference becomes negligible because the optimum scheduler tends to select users on an instantaneous basis who have both a strong direct link to their serving AP and {\em small} interfering links from surrounding APs. Interestingly, the minimization of the multicell interference term should take away some degrees of freedom in choosing the users with best direct links, however not sufficiently so to affect the overall capacity scaling.

Another interpretation of this result is in terms of our ability to find distributed scheduling schemes for maximizing the network capacity. The optimal multicell scheduler and power control solution would be hard to implement in practice. However from the observations above, a simple scheme based on each cell measuring the worst case SINR of each of its users (during e.g. a preamble) and selecting the users with the best worst case SINR as per (\ref{eq:SINRoptimal3}), will result in an quasi optimal behavior asymptotically (again, from a scaling perspective). Such a scheme does not require any exchange of information between the cells and the worst case SINR can be measured in one shot by each user and fedback to its serving AP.

These results come as a complement to previously reported findings \cite{kiani/oien/gesbert07,ebrahimi} which propose a near  optimal power allocation scheme, for fixed number of users, where
a fraction  of the transmitters are selected to be turned off while the rest operate at full power. It was observed experimentally \cite{kiani/oien/gesbert07} there that the fraction of off cells would go to zero when  the number of users grows large. Thus in a network with full reuse and greedy user scheduling, the optimal power control policy should be for all cells to operate at the power constraint.  The analysis of scaling of capacity provides a theoretical justification to this intuitive  result.

We now turn to a non symmetric network where users can experience different average SNR values depending on their position and conduct a similar analysis. However we will see that different capacity scaling rates are obtained compared with the symmetric network case.


\subsection{Capacity scaling with large $U$ in non-symmetric network}

We assume the path loss is determined by the user's distance to the emitting AP, both serving and interfering.
We consider a uniform distribution of the population in each cell. Thus $d_{u_n,n}$ (distance between user $u_n$ and its serving AP) is a random variable with non-uniform distribution $f_D(d)$. For a cell radius $R$, we find easily:

\be
f_D(d)=2d/R^2, \ \ d \in [0,R]
\ee
Further, the random process $d_{u_n,n}$ can be considered i.i.d. across users and cells, if users in each cell are dropped randomly in each disk\footnote{The considered coverage region can be assimilated with the inside area of each disk, in a disk-packing region of the 2D plane. Users dropped outside the disks can dropped from the analysis, as these will not affect the scaling law.}
Assuming $R=1$ for normalization, the distribution of $\gamma_{u_n,n}= \beta d_{u_n,i}^{-\epsilon}$  is given by (details omitted here):
\begin{equation}
f_{\gamma}(g) = \left\{ \begin{array}{ll}
\frac{2}{\epsilon} (\frac{g}{\beta})^{-\frac{2}{\epsilon}} \frac{1}{g} & \textrm{with $g \in [\beta, \infty )$}\\
0 & \textrm{with $g \notin [\beta, \infty)$}\\
\end{array} \right.
\label{eq:pathlossdist}
\end{equation}

 In order to get  upper  and lower bounds on performance, we are interested in the behavior of the following extreme values of product of independent random variables:
 \begin{eqnarray}
 \max_{u_n=1\hdots U} &  &\gamma_{u_n, n}  |h_{u_n,n}|^2 \ \ \mbox{for the interference-free case and} \nonumber \\
 \max_{u_n=1\hdots U} &  & \gamma_{u_n, n}  \omega_{u_n} \ \ \mbox{for the full-powered interference case} \nonumber
\end{eqnarray}
where $\omega_{u_n}$ is again defined as per (\ref{eq:omeg}).

\subsubsection{Extreme values of heavy-tail random variables}

The distribution of $\gamma_{u_n,n}$ shown in (\ref{eq:pathlossdist}) is remarkable in that it differs strongly from fast fading distributions, due to its {\em heavy tail} behavior. Tail behavior clearly plays a fundamental role in shaping the limiting distribution of the maximum value, hence also the scaling of capacity. Note that heavy tail is also observed in  {\em large scale} fading models such as log normal shadowing for instance. In order to study the extreme value of a product of random variables involving one heavy tailed variable, we need first to review the properties of so-called {\em regularly varying} random variables. See e.g. \cite{david03} for a definition of such variables, restated below:

 \begin{define}
A random variable $X$, with distribution (cdf) given by $F_X(x)$, is said to be regularly varying (at $\infty$) with exponent $-a$ if and only if:
\be
\frac{1-F_X(x)}{1-F_X(tx)}\rightarrow t^{a}  \ \ \mbox{when} \ \  t \rightarrow \infty
 \ee
 \end{define}

The lemma below shows how the definition above applies to our situation:

\begin{lemma}
Let $X=\gamma_{u_n,n}$. $X$ is regularly varying with exponent $-\frac{2}{\epsilon}$.
\label{lemama}
\end{lemma}

{\bf Proof:} A direct application of the definition above, with a distribution obtained from (\ref{eq:pathlossdist}):
\be
F_X(x)=1- \big( \frac{x}{\beta}\big)^{-\frac{2}{\epsilon}} \ \ \ x\geq \beta.
\label{pipi}
\ee
\vspace*{0.1cm}

An interesting aspect of regularly varying distributed random variable (R.V.) is that they are stable with respect to multiplication with other independent R.V. with finite moments as pointed out by the following theorem shown by Breiman \cite{breiman65}:

\begin{theorem} \label{theorem3}
 Let $X$  and $Y$ be two independent R.V. such that $X$ is regularly varying with exponent $-a$. Assuming $Y$ has finite moment $\mathbb{E}(Y^a)$, then the tail behavior of the product $Z=XY$ is governed by:
\begin{equation}
\label{eq:tailbehavior}
1-F_Z(z) \rightarrow \mathbb{E}(Y^a) (1-F_X(z)) \ \ \mbox{when} \ z \rightarrow \infty
\end{equation}
\end{theorem}

The idea behind this theorem is that when multiplying a  regularly varying R.V. with another one with finite moment, one obtains a heavy tailed R.V. whose tail behavior is similar to the first one, up to a scaling. In other words, heavy tail behavior tends to dominate over other distribution.

We now apply this result to $X=\gamma_{u_n,n}$ and $Y$ given by $Y=|h_{u_n,n}  |^2$ for the interference free case and $Y=\omega_{u_n}$ for the full-powered interference case, respectively.  Note that in both cases, $Y$ has finite moments.
The tail behavior of $Z=XY$ can then be characterized by the following lemma:

\begin{lemma} \label{lemma4}
 Let $X=\gamma_{u_n,n}$ be a R.V. with distribution given by (\ref{eq:pathlossdist}). Let $Y$ be an independent R.V. such that $\mathbb{E}(Y^{\frac{2}{\epsilon}})<\infty$. Then the tail of $Z=XY$ is governed by:
\begin{equation}
\label{eq:tailbehavior2}
1-F_Z(z) \rightarrow \mathbb{E}(Y^{\frac{2}{\epsilon}}) \left( \frac{\beta}{z} \right)^{\frac{2}{\epsilon}} \ \ \mbox{when} \ z \rightarrow \infty
\end{equation}
\end{lemma}

{\bf Proof:} A direct application of Theorem \ref{theorem3} using the distribution of $X$ shown in \ref{pipi}.

\vspace*{0.6cm}

The lemma above indicates that the tail behavior of the distribution of $X=\gamma_{u_n,n}$, characterized by Lemma \ref{lemama}, carries over to that of the product $Z=XY$.  As a consequence, $Z$ is also regularly varying with the same exponent $-\frac{2}{\epsilon}$.

We now complete our study by reviewing existing results on the extreme value of regularly varying R.V. Following \cite{david03}, a regularly varying variable can be classified to be of {\em Frechet} type. Extreme values of Frechet (or regularly varying) variables are characterized by use of the Gnedenko theorem, given in appendix I. For comparison, note that the random variables involved in the analysis of previous sections (Sec.\ref{sec:symm} and therein), belong to the so-called Gumbel category.
In our context, we have the following result:

\begin{lemma} \label{lemma5}
Let $Z_{u_n}=\gamma_{u_n,n} Y$ where $Y$ is a R.V. with finite moments, independent of $\gamma_{u_n,n}$. Then we have:
\be
\lim \mbox{Pr}\{ \max_{u_n=1 \hdots U} Z_{u_n} \leq \beta \mathbb{E}(Y^{\frac{2}{\epsilon}})^{\frac{\epsilon}{2}} U^{\frac{\epsilon}{2}}  t \} =e^{-t^{-\frac{2}{\epsilon}}} \ \ \forall t>0,  \ \ \mbox{when} \ U \rightarrow \infty
\ee
\end{lemma}

{\bf Proof:} We invoke Gnedenko's theorem \cite{gnedenko43} given in appendix I. It is easy to find that $a_U=\beta \mathbb{E}(Y^{\frac{2}{\epsilon}})^{\frac{\epsilon}{2}} U^{\frac{\epsilon}{2}}$ where $a_U$ is defined in the appendix.

\vspace*{0.6cm}

\subsubsection{Scaling law for interference-free case} \label{sec:intffree}

The inequality in (\ref{lemma5}) allows us to characterize the scaling law of capacity. Although a characterization in terms similar to those of previous section (i.e. finding a scaling law $l(U)$ for the SINR, such that the ratio of the SINR and $l(U)$ converges towards 1 when $U\rightarrow \infty$)  is possible when analyzing the capacity, such a task is very tedious and mathematically involved. For the sake of easing exposition, we somewhat loosen our definition of scaling of SINR below in a way that allows us to directly exploit Lemma \ref{lemma5} while preserving the key interpretations. Briefly speaking, we characterize the scaling of SINR {\em up to a multiplicative constant}. However it's important to note that the scaling of capacity itself, {\em within the standard definition of scaling}, will be determined precisely (thanks to the log function behavior).

The theorem below gives the  scaling law of SINR for the interference-free case in a non-symmetric network. First we give the following wider-sense definition of scaling:

\begin{define}
Let $g(U)$ and $l(U)$ two functions, defined over $U \in [U_0, +\infty]$, where $U_0$ is an arbitrary real. $g(U)$ is said to  {\em scale as} $l(U)$, i.e. $g(U)\sim l(U), \  \ U \rightarrow \infty $ when
\begin{eqnarray}{}
\mbox{Pr} (g(U) > v(U)) & \rightarrow 0, & \ \ \ \mbox{when} \ U \rightarrow \infty \nonumber \\
\mbox{Pr} (g(U) < w(U)) & \rightarrow 0, & \ \ \ \mbox{when} \ U \rightarrow \infty \nonumber \\
\end{eqnarray}

for any two functions $v(U)$ and  $w(U)$ such that $\frac{l(U)}{v(U)}\rightarrow 0$ and $\frac{w(U)}{l(U)}\rightarrow 0$, respectively.
\end{define}

Note that this notion of scaling can be interpreted as $g(U)$ grows neither significantly faster than $l(U)$, not does it grow significantly slower than $l(U)$.

 \begin{theorem} \label{theorem5}
 Let $h_{u_n,n}$, $u_n=1\hdots U$ be i.i.d. Gaussian distributed unit-variance random variables. Assuming  that $\gamma_{u_n,n}$ is i.i.d., distributed as per (\ref{eq:pathlossdist}), for $n=1\hdots N$.  Then for fixed $N$ and $U$ asymptotically large, the interference-free SNR scales like:
\begin{equation}
\label{eq:ggg}
\Gamma_n^{ub} \sim U^{\frac{\epsilon}{2}}
\end{equation}
\end{theorem}

{\bf Proof:}  Let $v(U)$ be any function growing faster than $U^{\frac{\epsilon}{2}}$, i.e. such that $\lim_{U\rightarrow \infty} U^{\frac{\epsilon}{2}}/v(U) =0$. Then let $t=v(U)/(\beta \mathbb{E}(Y^{\frac{2}{\epsilon}})^{\frac{\epsilon}{2}} U^{\frac{\epsilon}{2}})$. From Lemma \ref{lemma5} we have that
\be
 \mbox{Pr}\{ \max_{u_n=1 \hdots U} Z_{u_n} \leq v(U) \} \rightarrow \lim_{U\rightarrow \infty} e^{-t^{-\frac{2}{\epsilon}}}= 1
\ee
Equivalently, we have that $\mbox{Pr} (\max_{u_n=1 \hdots U} Z_{u_n} > v(U))  \rightarrow 0$. Similarly, we can prove  that any function $w(U)$ growing slower than $U^{\frac{\epsilon}{2}}$ will be such that $\mbox{Pr} \{\max_{u_n=1 \hdots U} Z_{u_n} < w(U)\}  \rightarrow 0$. Thus $\max_{u_n=1 \hdots U} Z_{u_n}$ scales as $U^{\frac{\epsilon}{2}}$ in the sense of definition 4.

\vspace*{0.3cm}

From the scaling of SNR above, we can infer the {\em exact} scaling on  the upper bound on capacity $\mathbb{E}( \mathcal{C}^{ub})$, as shown per the theorem below:

\begin{theorem} \label{theorem10}
 Let $h_{u_n,n}$, $u_n=1\hdots U$ be i.i.d. Gaussian distributed unit-variance random variables. Assuming  that $\gamma_{u_n,n}$ is i.i.d., distributed as per (\ref{eq:pathlossdist}), for $n=1\hdots N$.  Then for fixed $N$ and $U$ asymptotically large, the interference-free capacity scales like (i.e. the ratio of the two quantities converges to 1 almost surely):
\begin{equation}
\label{eq:capascaling5}
\mathbb{E}(\mathcal{C}^{ub}) \approx \frac{\epsilon}{2}  \log U  \ \ \mbox{for large $U$}
\end{equation}
\end{theorem}
\vspace*{0.6cm}

{\bf Proof:} See appendix II.

\vspace*{0.6cm}

We now proceed to determine the scaling laws in the case of full-powered interference.

\subsubsection{Scaling law for full-powered interference case}

We can derive the scaling laws  for the lower bound of SINR and capacity by following a strategy similar to Sec.\ref{sec:intffree}, simply by replacing the R.V. $|h_{u_n,n}  |^2$ by the R.V. $\omega_{u_n}$ which also has bounded moments. We obtain the following  result:

 \begin{theorem} \label{theorem6}
 Let $h_{u_n,i}$, $u_n=1\hdots U, i=1\hdots N$ be i.i.d. Gaussian distributed unit-variance random variables. Assuming  that $\gamma_{u_n,n}$ is i.i.d., distributed as per (\ref{eq:pathlossdist}), for $n=1\hdots N$.  Then for fixed $N$ and $U$ asymptotically large, the lower bound on SINR scales like:

\begin{equation}
\Gamma_n^{lb} \sim U^{\frac{\epsilon}{2}}
\end{equation}
 \end{theorem}

{\bf Proof:} We use the same proof as for Theorem \ref{theorem5},  with $X=\gamma_{u_n,n}$  but this time $Y=\omega_{u_n}$.

\vspace*{0.6cm}

Finally, from Theorem  \ref{theorem6}, we infer that the upper bound on capacity for a non-symmetric network exhibits an {\em exact} scaling defined as:

\begin{equation}
\label{eq:capascaling6}
\mathbb{E}(\mathcal{C}^{lb}) \approx \frac{\epsilon}{2}  \log U
\end{equation}

The proof for (\ref{eq:capascaling6}) is identical to that of Theorem \ref{theorem10} in Appendix II, but simply replacing $Y$ with $\omega_{u_n}$, which clearly does not change the scaling.

Remarkably, as in the case of the symmetric network, the results above (\ref{eq:capascaling5}) and (\ref{eq:capascaling6}) suggest that
multicell interference, no matter how strong, does not affect the scaling of the network capacity, if enough users exist {\em and} rate-optimal scheduling is applied. Furthermore, by virtue of the upper bound and lower bound exhibiting the same scaling law in (\ref{eq:capascaling5}) and (\ref{eq:capascaling6}) respectively,  the capacity under optimal scheduling and power allocation must behave like
\begin{equation}
\label{eq:capascaling7}
\mathcal{C}(\boldsymbol{U}^*,\boldsymbol{P}^*)  \approx \frac{\epsilon}{2}  \log U
\end{equation}

Two remarks are in order. First, in the symmetric network case, a suboptimal but fully distributed resource allocation based on constant (full) power transmission at all transmitters and scheduling policy based on  (\ref{eq:SINRoptimal2}) will actually result in the best possible  scaling law of capacity for the network. Second, we observe that we obtain a much greater rate growth than in the case of the symmetric network. This is due to the amplified multiuser diversity gain due to the presence of unequal path loss across the user locations in the cell. This results from a scheduler which, in a quite unfair fashion admittedly, tends to select users closer to the access point as more users are added to the network.

\subsection{Discussion on channel models and exclusion area around the AP}
Interestingly, the theory on regularly varying variables stipulates that multiplication of the path loss variables by any small scale fading variable with finite moments will preserve its heavy tail behavior. This means that our result shown in
(\ref{eq:capascaling7}) is in fact valid for a wider class of fading channel models, such as Nakagami, Rice, etc.
On a different note, one may wonder how close users can be assumed to get to the access point in practice. Let us imagine that
a small disk of exclusion, with the AP at its center, prevents users to getting to close to the AP. As a by product, the disk also serves the purpose of maintaining the validity of the path loss model, which
 may  not be reasonable in the close vicinity of the AP.
 In this case, one may expect two successive regimes for the capacity scaling as $U$ grows. In the first regime, when the number of users is still moderate, the scaling is dominated by the path loss effect, with a behavior such as shown in  (\ref{eq:capascaling7}). In the second regime, when enough users are already accumulated near the exclusion circle, it is the turn of the tail behavior of small scale fading to dominate and the scaling will be characterized by (\ref{eq:capascaling22}). This situation is investigated in the next section.

As the growth would be ultimately limited by that the tail of the small-scale random fading in practical situations, one may also wonder how accurately Chi-square distributions model reality in real-world wireless channels. Clearly, this discussion is inherent in all previous studies dealing with scaling laws and asymptotic performance analysis. Nevertheless it is important to keep in mind the basic law of power preservation which indicates that no matter how many users are considered, the most favorable users can't receive more power than what was actually transmitted.  This simple fact will impose a hard limit on the SNR which in turn limits the domain of validity of our scaling in terms of the number of users $U$. Although we believe a specific analysis of the validity domain will rely on yet unexpored channel model properties (tail properties of the pdf are less explored than the behavior near zero which characterize outage) and is outside the scope of this paper, it remains clear that this domain is wide enough for the analysis to be meaningful since the power preservation limit is reached only when the small scale fading is in the order of the inverse of path loss, which would require  very large fading coefficients in practice (several tens of dB).

\section{Numerical evaluation}
\begin{figure}[h]
\centering
\includegraphics[scale=0.85]{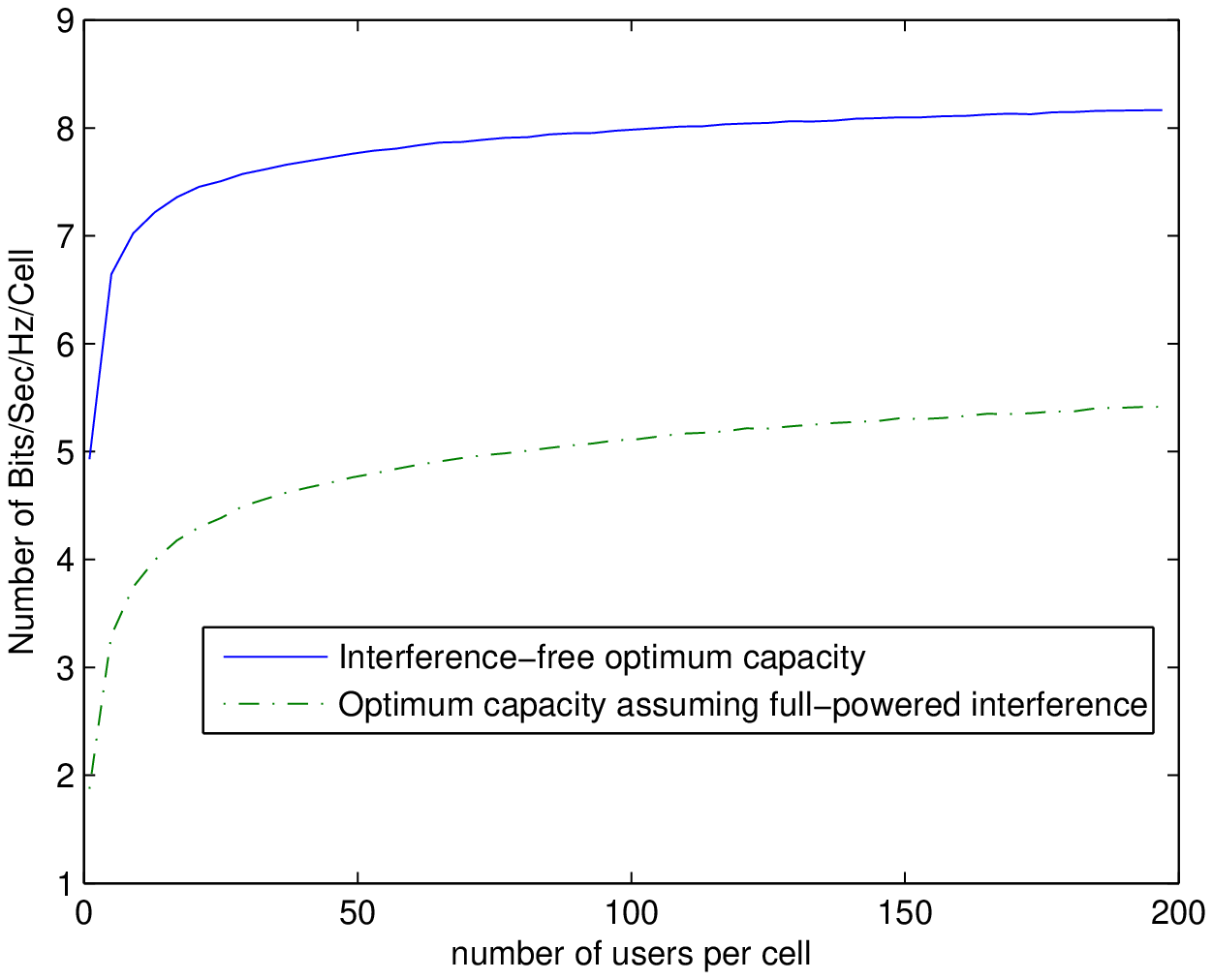}
\caption{Scaling of upper and lower bounds of capacity versus $U$ for a symmetric network ($N=4$). The observed scaling for both curves is in $\log \log U$. }
\label{fig:sym}
\end{figure}

\begin{figure}[h]
\centering
\includegraphics[scale=0.85]{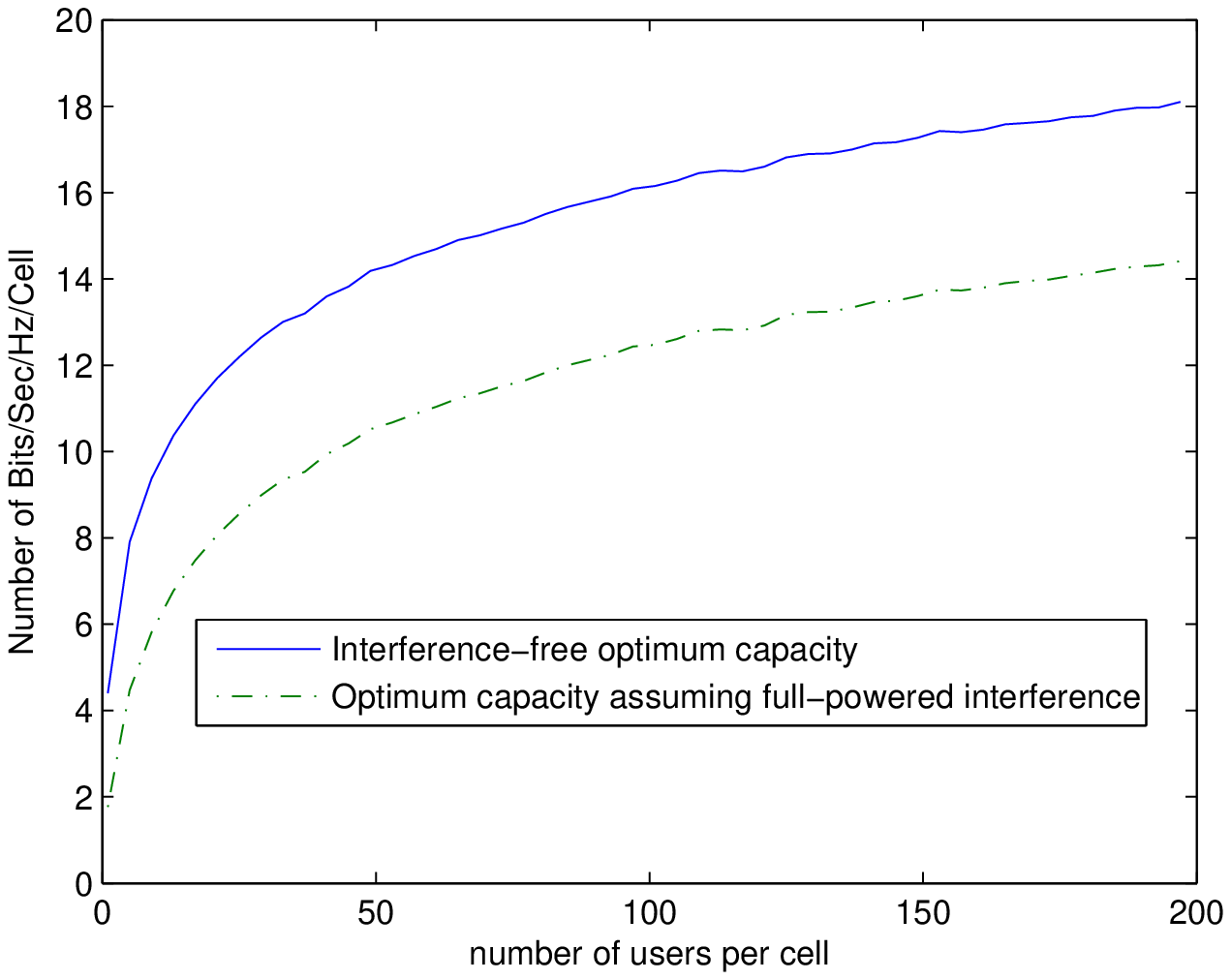}
\caption{Scaling of upper and lower bounds of capacity versus $U$ for a non-symmetric network (unequal average SNR) ($N=4$). The observed scaling for both curves is in $\log  U$.}
\label{fig:nonsym}
\end{figure}

\begin{figure}[h]
\centering
\includegraphics[scale=0.85]{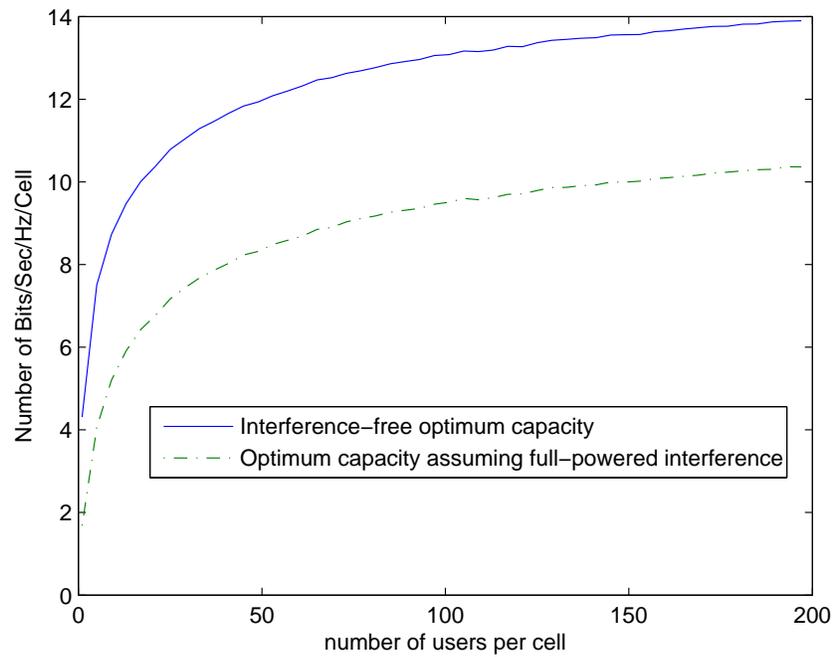}
\caption{Scaling of upper and lower bounds of capacity versus $U$ for a non-symmetric network with a an exclusion disk around each AP of radius 0.1. ($N=4$).The observed scaling in the transitory regime  is in $\log   U$, then in $\log   \log U$ in the asymptotic regime.}
\label{fig:hybrid}
\end{figure}

We validate the asymptotic behavior of the multicell sum rate when $U$ grows large with Monte Carlo simulations. We use a  network with $N=4$ cells,  unit cell radius and the following parameters $\beta=1/16$, $\epsilon=4$, $P_{max}=1$, $\sigma^2=0.02$. I.i.d. flat Rayleigh fading is considered in addition to the path loss based power decay. We consider three scenarios for user location, as mentioned previously in this paper. First, we consider cells with  users located on a circle with distance 0.5 away from the AP (symmetric network). Then we consider a non-symmetric distribution of average SNR by drawing users randomly in the cell. Finally we consider an hybrid scenario where users are drawn unform randomly over the cell but kept outside an exclusion disk of radius 0.1 around the AP.     In all cases, we evaluate the upper and lower bound on per-cell data rates (see Fig.\ref{fig:sym}, Fig.\ref{fig:nonsym}, Fig.\ref{fig:hybrid} and observe the identical rate growth of the lower and upper rate bounds. This also shows  that the capacity obtained with exhaustive user and power level selection also has the same growth rate. The observed rate growth in $\log \log U$ for the symmetric network and in $\log U$ for the non symmetric network confirms our earlier theoretical claims.
In Fig.\ref{fig:hybrid}, we observe a scaling behavior with two distinct regimes with a $\log U$ in the moderate number of users $U$ and $\log \log U$ for high number of users, thus confirming our intuition for what could happen in a realistic network.

\section{Conclusions}
We present an extreme value theoretic analysis of network capacity for maximum sum rate multicell power allocation and user scheduling.
We derive scaling laws of capacity when the number of users per cell grows large, both in cases where the users have same average SNR and path loss dependent SNR. We show that in both cases, 1-the effect of intercell interference on rate scaling tends to be negligible asymptotically, and 2-should intercell interference be considered, an asymptotically optimal allocation procedure is given based on full power allocation at all transmitters, which is furthermore completely distributed.  We show that the growth rate of capacity is exponentially faster in the case of a system with unequal distance-based average SNR.

\newpage

\appendices

\section{}
The following theorem is due to Gnedenko \cite{gnedenko43} (1943) and states the following property for regularly varying distributions:
\begin{theorem}
Let $Z_i$ an i.i.d. random process. Then  $Z_i$ has a regularly varying distribution with exponent $a$ if and only if
\begin{equation}
\label{eq:tailbehavior22}
\lim \mbox{Pr}\{ \max_{i=1\hdots U} Z_i \leq a_U t \} =e^{-t^{-a}}  \ \ \forall t>0 \ \ \mbox{when} \ U \rightarrow \infty
\end{equation}
where $a_U$ is a sequence such that $1-F_{Z}(a_U)=\frac{1}{U}$.
\end{theorem}

\section{}
From Lemma \ref{lemma5} we have that
\be
\label{eq:tailbehavior222}
\lim \mbox{Pr}\{  \Gamma_n^{ub} \leq \beta \mathbb{E}(Y^{\frac{2}{\epsilon}})^{\frac{\epsilon}{2}} U^{\frac{\epsilon}{2}}  t \} =e^{-t^{-\frac{2}{\epsilon}}} \ \ \forall t>0,  \ \ \mbox{when} \ U \rightarrow \infty
\ee
Since the SNR $\Gamma_n^{ub}$ is growing large in each cell by virtue of Theorem \ref{theorem5}, the capacity can be approximated by:
\begin{equation}
\mathcal{C}^{ub} \approx \frac{1}{N}\sum_{n=1}^{N}\log
\Gamma_n^{ub}  .
\label{eq:upperbound22}
\end{equation}
when $U$ grows large. From (\ref{eq:tailbehavior222}), we write
\be
\lim \mbox{Pr}\{ \log \Gamma_n^{ub} \leq \log (\beta \mathbb{E}(Y^{\frac{2}{\epsilon}})^{\frac{\epsilon}{2}}) + \log t + \frac{\epsilon}{2} \log U  \} =e^{-t^{-\frac{2}{\epsilon}}} \ \ \forall t>0,  \ \ \mbox{when} \ U \rightarrow \infty
\ee
Now, taking $t=\log U$ we infer that
\be
\log \Gamma_n^{ub} \leq \log (\beta \mathbb{E}(Y^{\frac{2}{\epsilon}})^{\frac{\epsilon}{2}}) + \log \log U + \frac{\epsilon}{2} \log U   \ \ \mbox{almost surely when} \ \ \ U \rightarrow \infty
\label{pipi2}
\ee
On the other hand, taking $t=1/\log U$, we obtain that
\be
\log \Gamma_n^{ub} \geq   \frac{\epsilon}{2} \log U  - \log \log U \} \ \ \mbox{almost surely when} \ \ \ U \rightarrow \infty
\label{pipi3}
\ee
From (\ref{pipi2}) and (\ref{pipi3}), we conclude that
\be
\frac{\log \Gamma_n^{ub}}{\frac{\epsilon}{2} \log U} \rightarrow 1  \ \ \mbox{almost surely when} \ \ \ U \rightarrow \infty
\ee
From the isotropy of the network, this shows that $\mathcal{C}^{ub}$ (and {\em a fortiori} its average) scales as $\frac{\epsilon}{2} \log U$.


\end{document}